# Universal Wireless Power Transfer for Energy Security, Availability and Convenience

Dinh Hoa Nguyen[1,2,*], Andrew Chapman[1],

[1]International Institute for Carbon-Neutral Energy Research (WPI-I2CNER)

[2]Institute of Mathematics for Industry (IMI)

Kyushu University, Fukuoka 819-0395, Japan

**Abstract**—This article proposes a novel system concept named universal wireless power transfer, in which power can be wirelessly transferred between different entities (e.g. vehicles, robots, homes, grid facilities, consumer electronic devices, etc.) equipped with proper energy transmitters and receivers, whether stationary or in motion. This concept generalizes individually existing wireless power transfer systems, where a specific wireless power transfer technology is used, and where the wireless energy transmitter or receiver is fixed. As a result, energy mobility, flexibility, and convenience are significantly improved by the proposed universal wireless power transfer concept in this study. Moreover, factors relevant to system energy efficiency are analyzed according to each utilized wireless power transfer technology. Necessary market mechanisms for such a concept to be successfully deployed are also introduced, along with an analysis of the benefits engendered in terms of improving energy systems, the environment, human comfort, and convenience. Finally, a discussion of the proposed concept, policy implications and recommendations for future research directions which will underpin universal wireless power transfer systems are presented.

## 1. Introduction

There is increasing worldwide concern regarding greenhouse gas (GHG) emissions due to their impacts on climate change and society at large. Among energy sectors, transportation and power grids are significant contributors to GHG emissions, requiring decarbonization as a priority [70]. The combination of two emerging technology regimes, renewable power generation and transportation electrification, offer practical and promising solutions, where the former provides a source of clean energy for the latter which in turn actively contributes to the additional deployment of renewable energy (RE) for energy generation, storage and proactive use. For instance, solar energy can be utilized for both solar-powered charging stations and direct solar charging for electrified vehicles (EVs) and for hydrogen creation for fuel cell vehicles (FCVs) through the electrolysis process [1].

However, massive deployment of EVs (including FCVs) is currently hampered due to several barriers: (i) high cost; (ii) limited range due to limited battery or fuel cell (FC) capacity; and (iii) a limited number of charging points. In addition, other factors of the local transportation and grid infrastructure can also hinder the popularity of EVs. Recently, flying EV (FEV) are emerging as a promising and interesting alternative for solving ground transportation problems, gaining attention worldwide from both major automobile makers and start-up companies. The potential advantages of flying vehicles over ground vehicles are due to the non-existence of traffic lights and traffic jams in the sky. Hence, shorter traveling time and lower energy usage can be achieved. As such, many countries are considering the deployment of flying vehicles for passenger use, including the USA, New Zealand, Dubai, and Singapore [68]. On the commercial side, companies such as Cartivator [21] and NEC [22] have demonstrated prototypes of flying cars and personal vertical takeoff and landing (VTOL) vehicles, and a roadmap has been announced in Japan for their future deployment [20]. The number of companies working on personal flying vehicle and flying taxi projects is also rapidly increasing including major corporations such as Boeing [23], Audi and Airbus [24], Google [25], Uber [26], and BMW-supported FC Skai [27]. A more comprehensive summary of current FEV companies is available in [28] and [29]. Nevertheless, these FEVs suffer the same mentioned barriers as ground EVs.

*Corresponding author. Emails: hoa.nd@i2cner.kyushu-u.ac.jp, chapman@i2cner.kyushu-u.ac.jp

It is noteworthy that the recent development of (manned and unmanned) aerial vehicles as well as ground robot delivery services also presents challenges for the necessary energy management and charging infrastructure, similar to that for EVs. To extend travelling range, on-the-fly charging capabilities for vehicles and robots becomes a necessity. On the other hand, vehicle users have strong preferences toward energy readiness and convenience, in addition to low-carbon energy sources. Hence, energy ultra-mobility, which is available anywhere, anytime, is necessary for a connected, high quality life, in which vehicles (whether ground, aerial or maritime) and robots play an important role.

To resolve the above sociotechnical challenges, an attractive solution can be proposed based on wireless power transfer (WPT) and RE sources. As its name indicates, WPT means power is transferred from one object to another without being connected by any cable, thereby eliminating the inflexibility, inconvenience, and technical concerns of wired charging. Thus, WPT has gained much attention from both industrial and academic communities recently. In the literature, a number of reviews have been performed for inductive WPT (IWPT) detailing charging pads and powered electronics, e.g., pad shapes, coil designs, compensation topologies, etc., in [7], [10], [19] for EVs; for simultaneous wireless information and power transfer from a signal processing point of view [33]; for unmanned aerial vehicle (UAV) communication [3], for UAV wireless charging using power transmission lines [49], and for different UAV wireless charging technologies [50]. Other review articles e.g. [64], [65] evaluated specific use cases such as magnetic resonance for EV charging [64] and WPT of renewable energy. While [64] summarizes core technologies for resonant IWPT for EVs and compares academic and industrial research, [65] focuses on the use of solar energy for WPT. More details on WPT technologies for ground, aerial, and maritime use cases will be outlined in Section 3.

Under the current regime, for all sectors of transportation, logistics, wearables and consumer electronics and biomedical healthcare, etc., state-of-the-art available WPT technologies have been developed in such a way that wireless charging requires at least one fixed (stationary) object, usually the wireless energy transmitter, while the wireless energy receiver is able to be moving. For example, in current EV wireless charging systems, the charging coils are fixed at charging stations, grid charging points, or under designated wireless charging lanes [10], [19]. Similarly, wearable and consumer electronic devices must be put on top of, or near a fixed wireless charging base. Such requirements of fixed object dependent charging do not address the issues of mobility, flexibility, availability, and convenience of wireless charging.

Another recent research direction is focused on power transfer between EVs, however, energy is transferred via wires installed in designated parking slots [30–32]. Very few works have investigated WPT between moving vehicles. Recent studies such as [3], [33–36] analyze WPT between moving aerial vehicles by means of radio frequency (RF), laser wireless powering, IWPT or capacitive wireless power transfer (CWPT); respectively. For an up-to-date review of laser WPT, readers are referred to [37]. Optical wireless power transfer (OWPT) between moving objects has recently been reported in [72]. With regard to terrestrial EVs, automakers are moving toward vehicles which incorporate solar cells, enabling these vehicles to self-charge, and also enhance the ability of energy transfer between vehicles [39], [40], [41].

In order to overcome the above-mentioned drawbacks of current WPT technologies, this research proposes a systemic concept of universal WPT for all energy sectors in which wireless energy transmitters and receivers can be in motion. Therefore, anytime, anywhere availability of energy as well as human convenience can be achieved. Market mechanisms for enabling WPT energy transactions are also proposed based on distributed ledger technologies. For conciseness, the proposed concept of universal WPT will focus on EVs, but the same concept is applicable to other systems such as robotics, consumer electronics, logistics, biomedical healthcare, and the internet of things (IoT), etc.

The remainder of this paper is organized as follows. Section 2 briefly reviews the advantages and disadvantages of current WPT systems for EVs. Details of the proposed universal WPT system are provided in Section 3, in which system energy efficiency is presented for each WPT technology alongside market mechanisms to realize the proposed system. Section 4 analyzes the positive impacts of the proposed universal WPT system from various aspects. Finally, Section 5 summarizes the paper, identifies the limitations of the proposed concept, and provides directions for further improvements in future research efforts.

## 2. Advantages and Disadvantages of State-of-the-Art Vehicle WPT Systems

### 2.1. Ground Vehicles

As outlined in the introduction, several review articles, e.g. [7], [10], [19], have been detailed for WPT systems for ground EVs along with specifics of applicable technologies such as IWPT, resonant IWPT, and CWPT. These reviews are not duplicated here, instead the present study briefly summarizes the advantages and disadvantages of those WPT systems.

Different from aerial and underwater vehicles, vehicles moving on the ground have physical roads which are potentially beneficial in the context of WPT. A pertinent example is the case of WPT transmitters embedded under road lanes, while energy receivers are attached under the EV chassis. Thus, whenever EVs pass under-road transmitters, power is wirelessly sent to them. This type of WPT is often referred to as dynamic or in-motion charging in the literature. Though the construction and maintenance costs of such wireless charging lanes are currently relatively high, they are attractive to EV proponents as they can enable mass deployment of EVs [71]. Further, RE sources such as solar panels and wind turbines can be located along roadsides to supply clean energy for wireless charging lanes, underpinning carbon-free WPT systems.

An additional current WPT system for EVs, referred to as static charging, utilizes fixed charging stations located at homes, offices, or shopping centers, etc. Unlike dynamic charging, static charging is applicable not only for ground EVs but also for aerial and underwater EVs. At present, static WPT is deployed more widely than dynamic or in-motion charging, however static WPT requires EVs to park, reducing the mobility and convenience of EV users. In addition, limited numbers of charging points can also impinge upon user comfort and convenience.

## 2.2. Aerial Vehicles

Current charging technologies for aerial vehicles such as UAVs include battery swapping, FC swapping, and WPT at landing pads [50]. Although these approaches incur no or low energy losses, they require UAVs to land in order to charge, reducing flexibility and convenience, and eliminate the possibility for uninterrupted flight. In addition, swapping-based approaches lead to limitations on the number of batteries and FCs available for substitution, increasing system costs and burdens on available resources. The landing pad WPT approach, based on resonant IWPT or CWPT, similar to ground vehicle approaches is likely to be preferred considering trade-offs between flexibility, convenience, and cost. A recent review [49] has specifically concentrated on resonant IWPT for UAVs using high-voltage power transmission lines, a promising solution for massively deployed UAV wireless charging requirements. Future charging technologies being considered for UAVS include approaches such as those utilizing microwaves, radio frequency (RF) waves or laser [3] [50], or ultrasonic approaches, however, low system efficiency is a significant drawback for each of these approaches. In addition to wireless charging approaches, a tethered charging system for UAVs has also been proposed, not without its own safety challenges [3] [50].

## 2.3. Underwater Vehicles

Current system configurations for underwater vehicle wireless charging are broadly divided into docking and docking-free approaches. For docking approaches, the basic system setup includes a cage [52] [60], or a base [53] [61] in which the transmitting coil is located, and underwater vehicles need to be docked to such a cage or base such that the receiving coil in the vehicle can be aligned with the transmitting coil. On the other hand, in docking free approaches, underwater vehicles do not need to be docked, but are controlled in order to stay in close proximity to underwater charging stations, such as the floating station proposed in [55]. Currently docking-free approaches are not as popular as docking methods, however in the near future these approaches may have an advantage due to their convenience and comparatively simple mechanical design.

The employed technologies for underwater vehicle wireless charging are also diverse, including IWPT, CWPT, laser, LED arrays, RF, and ultrasonic approaches. The IWPT technology were appraised in [51-54] for docking methods, while RF was utilized in a docking-free approach in [55]. Meanwhile, a CWPT design for freshwater use was proposed in [56] to clarify the behaviors of the coupling coefficient and its product with the Q-factor in the coupler. In another CWPT study [57], bidirectional WPT was studied for seawater using two H-bridge circuits. Separately, light can be used as a means of WPT for underwater vehicles, for example in the form of blue LED arrays in [58] or laser diodes in [9]. Even though WPT utilizing light has a lower efficiency than IWPT and CWPT technologies, it is remarkable in that it can transfer energy over long distances and eliminate the inherent IWPT and CWPT disadvantage of complex electronics. Finally, ultrasonic waves have been employed as a WPT technology for battery-less underwater sensor nodes in [59] toward the development of an internet of underwater things. Additional research on the power systems and energy conversion methods for autonomous underwater vehicles has been undertaken in [63].

## 3. Proposed Universal WPT Regime

This section details a universal WPT regime, describing grid and charging infrastructures, as well as the market mechanisms which will underpin the proposed concept. Cognizant of gaps or shortcomings in existing technologies, unique properties are incorporated to ensure that energy necessary for operation is available at all times.

### 3.1. System Description

To overcome the drawbacks of currently used wireless charging technologies as reviewed previously, we propose a novel concept of 'universal WPT', in which energy can be wirelessly transferred between any two objects, whether stationary or in-motion, located on the ground, in the air, on the water surface, or underwater, via any available WPT technologies (IWPT, resonant IWPT, CWPT, OWPT, etc.).

For clarity of illustration, the proposed universal WPT concept is first presented for EVs, where power is wirelessly transferred between EVs, or between an EV and a charging lane/base/station, as depicted in Figure 1. EVs are equipped with hybrid energy storage technologies, incorporating both batteries and supercapacitors as it is well-known that popular battery types, such as Li-ion have high energy density but suffer from low power density, meaning that it takes a long time for them to charge or discharge. In contrast, supercapacitors have high power density but low energy density, therefore they can be promptly charged or discharged but cannot store as much energy as their Li-ion counterparts. The complementary combination of batteries and supercapacitors offers significant advantages for the proposed universal WPT concept, in which supercapacitors are used to rapidly charge or discharge EVs through WPT, while batteries are employed as larger scale storage systems to drive EVs over long distances.

Another appealing feature of the proposed universal WPT concept is that energy transfer can be wirelessly made between multiple objects, and is not limited to only two objects. For instance, one EV can be simultaneously charged by several other EVs, and vice versa, several EVs can be charged at the same time by one EV.

Based on currently available WPT technologies, OWPT and microwave WPT appear most appropriate for the proposed WPT between always-moving entities, or between moving receivers and one fixed transmitter at long distances, while other WPT technologies can be used for wirelessly transmitting power in other contexts.

Under the proposed system, the current issue of limited range and charging infrastructure available for EVs can be overcome through on-the-fly charging, extending vehicle range without the need for fixed charging stations or battery and FC swapping. Whenever an EV lacks sufficient energy for a trip, its energy demand will be uploaded to a registered information and communication network such that network connected agents (which can be a ground, aerial, water-surface, or underwater vehicles depending on the specific situation) can fulfil these energy needs. Such a system meets the requirements for anywhere, anytime availability of energy, enhancing human convenience and comfort.

Similar universal WPT systems are also applicable to wearable and consumer electronics (e.g. smart phones, laptops, tablets, smart watches, etc.), robotics, biomedical healthcare (e.g., implant devices, wearable sensors, etc.), internet of things (IoT) devices, and so on.

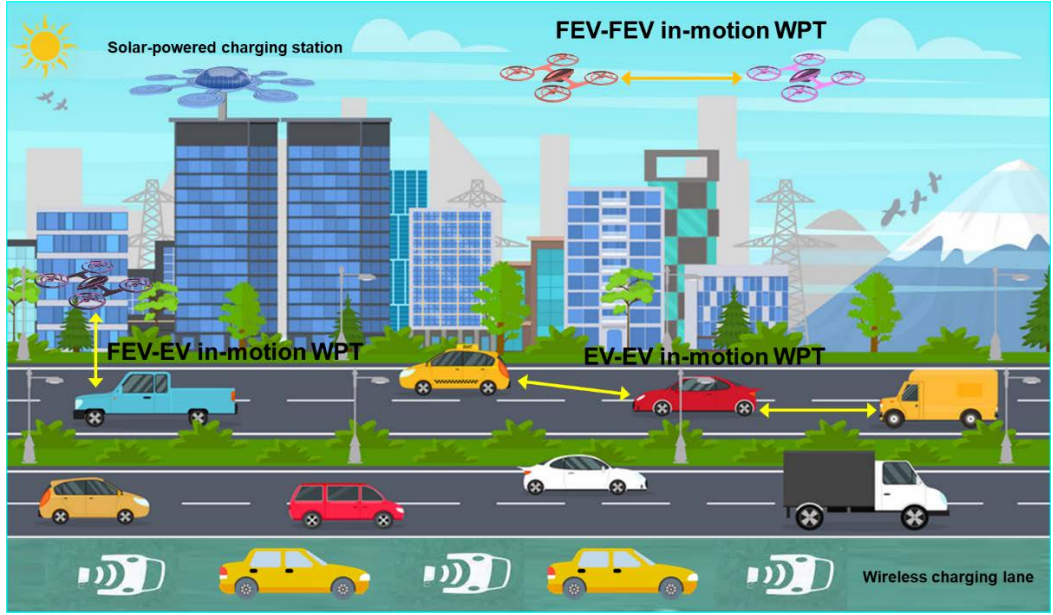

a. Ground and aerial EVs

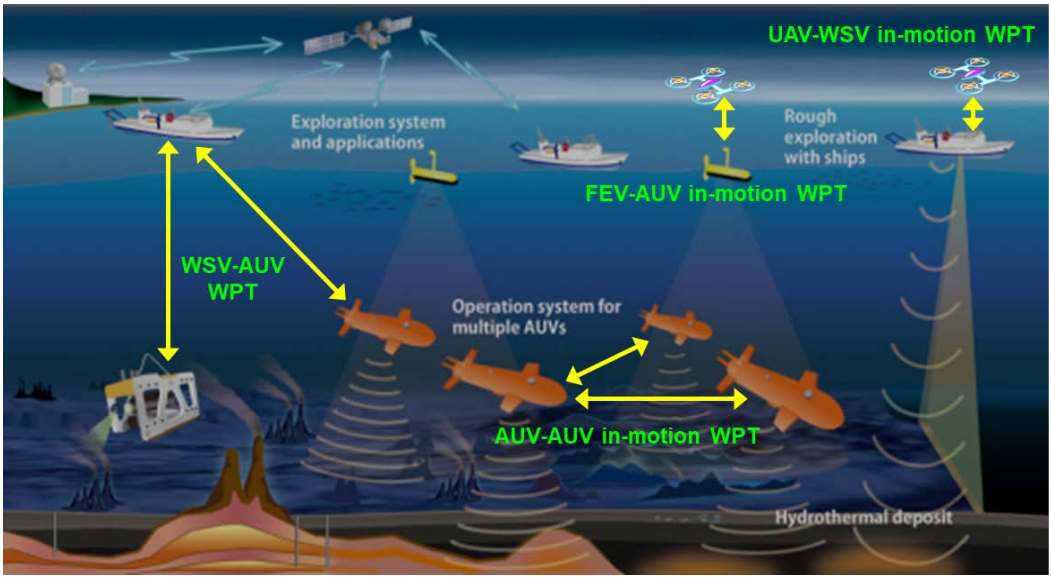

b. Maritime and aerial EVs (UAV: underwater autonomous vehicle, WSV: water-surface vehicle)

Figure 1. Visionary image for the proposed concept of universal wireless power transfer

### 3.2. System Energy Efficiency

As for any WPT system, energy efficiency is an important issue to be considered for the proposed universal WPT system. Herein, the overall system energy efficiency $\eta_{wpt}$ is calculated as in Eq. (1),

$$\eta_{wpt} = \eta_{tr} \times \eta_{env} \times \eta_{rc} \qquad (1)$$

where $\eta_{tr}, \eta_{env}, \eta_{rc}$ represent the efficiency of the wireless energy transmitter, of the wireless energy transfer through the environment, and of the wireless energy receiver; respectively. Each of the above energy efficiencies depends on which specific WPT technology is utilized. However, in general $\eta_{wpt}$ is inversely proportional to the distance between the energy transmitter and energy receiver. Another way for deriving $\eta_{wpt}$ is to directly compute the ratio between the energy obtained at the receiver and that consumed at the transmitter.

It is worth emphasizing that energy efficiency of the overall system is affected by the alignment between the transmitter and the receiver. In other words, it is crucial to maintain line-of-sight between the energy transmitter and receiver for obtaining high energy efficiency. With IWPT, resonant IWPT, and CWPT technologies, the wireless energy transmitter and receiver should be properly aligned both vertically and horizontally, because positions and directions of the coils (in IWPT and resonant IWPT technologies) or metal plates (in CWPT technology) are usually inflexible. Nevertheless, in OWPT technology, additional control systems can be added to the wireless energy transmitter so that it can be turned around to track the energy receiver, hence a line-of-sight can be adaptively created. This point becomes essential when both the wireless energy transmitter and receiver are moving. Note also that a suitable inter-object distance is a requirement for high system energy efficiency, especially when all objects are in motion. In the case of EVs (whether ground, aerial, water-surface, or underwater EVs), such distance should be greater than a safe distance for avoiding collisions.

### 3.2.1. Transmission efficiency in OWPT systems

In an OWPT system using light emitting diodes (LEDs), $\eta_{env}$ can be computed as follows. The transmitted optical power (or radiant flux) from an LED is calculated by (see, e.g. [84]),

$$P_{tr} = \int_{\lambda_{min}}^{\lambda_{max}} \int_0^{2\pi} \Phi_e(\theta, \lambda) d\theta d\lambda \qquad (2)$$

which is the integral of the energy flux $\Phi_e$ in all directions. In equation (2), $\lambda_{min}$ and $\lambda_{max}$ are minimum and maximum wavelengths which can be detected by the optical receiver. Next, the received optical power at the receiver surface, assumed direct line of sight transmission, is computed as follows [84],

$$P_{rc} = H(0)P_{tr} \qquad (3)$$

where $H(0)$ is the DC gain of the optical transmission link calculated by

$$H(0) = \begin{cases} \frac{(m_l+1)A}{2\pi d^2} \cos^{m_l} \phi \, g_f(\varphi) g_c(\varphi) \cos \varphi : 0 \leq \varphi \leq \varphi_w \\ 0 : \varphi > \varphi_w \end{cases} \qquad (4)$$

in which $A$ is the physical area of the optical receiver, $d$ is the optical transmission distance, $\varphi$ is the angle of incidence (i.e. the angle at which the receiver sees the transmitter), $\phi$ is the angle of irradiance (i.e. the angle at which the transmitter sees the receiver), $g_f(\varphi)$ is the gain of an optical filter if exists, $g_c(\varphi)$ is the gain if an optical concentrator if exists, and $\varphi_w$ is the width of the field of view (FOV) at the optical receiver. Moreover, $m_l$ denotes the order of Lambertian emission which is given by the semi-angle at half illuminance $\phi_{1/2}$ of an LED as follows,

$$m_l = -\frac{\ln 2}{\ln(\cos \phi_{1/2})} \qquad (5)$$

It is obvious from (4) that the transmitted optical power is inversely quadratically decreased with the transmission distance, when the angle of incidence is not greater than the angle of irradiance. Moreover, $H(0)$ can be regarded as the transmission efficiency through the environment $\eta_{env}$.

For laser-based OWPT systems, the flux received by the receiver from a laser source is calculated utilizing [37], as follows:

$$\phi = \frac{R_s A_s \eta_{abs}}{d^2} \tag{6}$$

where $R_s$ is the radiance at the source, $A_s$ is the total area of the beam source, and $d$ is the transmitting distance, and $\eta_{abs}$ represents the efficiency as the laser beam is absorbed by the environment. As seen in (6), the transmission efficiency of laser-based OWPT is also decreased quadratically with transmitting distance increase, similar to the LED-based OWPT.

### 3.2.3. Transmission efficiency in microwave WPT systems

In radio frequency (RF) based WPT systems, the transmitting efficiency is captured by the Friis transmission equation (see, e.g. [86]), as follows:

$$\eta_{env,max} = \frac{A_r A_t}{d^2 \lambda^2} \tag{7}$$

where $d$ is the transmitting distance, $A_r$ and $A_t$ are effective aperture of the receiving and transmitting antennas, respectively, and $\lambda$ is the wavelength.

### 3.2.3. Transmission efficiency in IWPT systems

For IWPT and resonant IWPT, transmitting efficiency is calculated as follows [19]:

$$\eta_{env} = \frac{R_L}{\frac{(R_2+R_L)^2}{k^2 Q_1 Q_2 R_2} + R_2 + R_L} \tag{8}$$

where $Q_1$ and $Q_2$ are quality factors of the primary (transmitter) and secondary (receiver) sides determined as $Q_i = \frac{\omega L_i}{R_i}, i = 1,2$, and the coupling coefficient is defined by $k = \frac{L_m}{\sqrt{L_1 L_2}}$, in which $L_1$ and $L_2$ are the inductance at the primary and secondary coils, $L_m$ is the mutual inductance between primary and secondary coils, and $R_L$ is the load resistance.

### 3.2.4. Transmission efficiency in CWPT systems

In CWPT systems, transmitting efficiency depends on circuit designs, which may have the following form. For parallel CWPT systems, the following expression of transmitting efficiency was derived in [56]:

$$\eta_{env,max} = \frac{\sqrt{1+(kQ)^2}-1}{\sqrt{1+(kQ)^2}+1} \tag{9}$$

where $\eta_{env,max}$ is the maximum achievable transmitting efficiency, $k$ is the coupling coefficient, $Q$ is the quality factor.

For singular CWPT systems, the maximum transmitting efficiency can be computed by [85],

$$\eta_{env,max} = \frac{k^2 Q_1 Q_2}{\left(\sqrt{1+k^2 Q_1 Q_2}+1\right)^2} \tag{10}$$

where $Q_1$ and $Q_2$ are the quality factors of the primary and secondary circuits.

### 3.3. Market Mechanisms

To enable the proposed universal WPT approach, market mechanisms need to be set up to facilitate energy sharing and trading between EVs or between EVs and facilities such as power grids, buildings, etc. Multiple solutions are available for designing such market mechanisms, including wireless energy service providers, peer-to-peer (P2P) energy trading, etc., and energy transactions can be secured through blockchain technologies. Market mechanisms can be summarized as the 'Distributed Ledger Energy Market', as demonstrated in Figure 2.

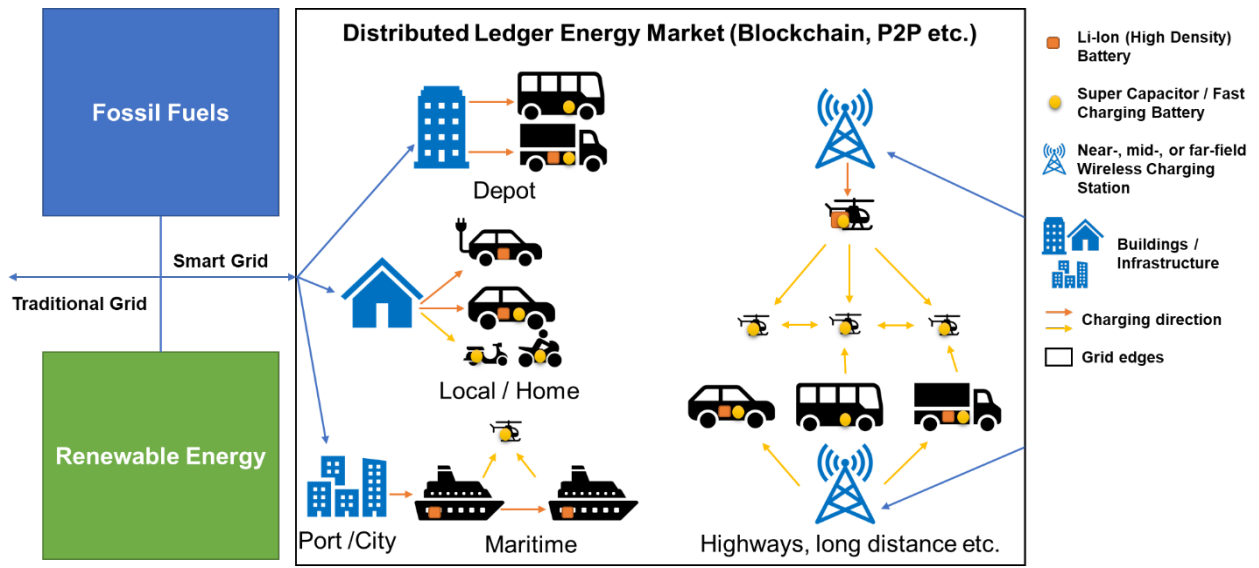

Figure 2. Market demonstration for the proposed concept

To participate in distributed ledger energy markets, EVs need to be equipped with an information and communication system that can register and execute energy transactions. Once an energy contract or bid is successfully negotiated, the WPT will be physically undertaken.

The system design proposed in Figure 2 can be neatly summarized according to the traditional grid on the left-hand side, made up of the existing generation plant, both fossil and RE-based, and the right-hand side, which describes a connected yet discrete distribution system. For the left-hand side, the increase of renewable energy is the main goal, with the distribution system underpinning this goal through the management of intermittent renewable sources by storing energy across the fleet of vehicles, in addition to that used in the traditional grid. It is anticipated that the market will ensure lowest cost delivery of electricity to vehicles within the system, as it prioritizes (or at least encourages) the use of renewable energy sources which have the strongest technological learning curves and the lowest marginal cost among generation options. The proposed concept also takes advantage of fit-for-purpose battery types, engaging Lithium-ion batteries for long-distance applications, and supercapacitors or fast charging varieties for always connected or short-distance applications.

## 4. Positive Impacts of Proposed Universal WPT Regime

### 4.1. Benefits for Energy Systems

EVs, together with the proposed universal WPT concept, will significantly increase the flexibility and mobility of energy grids in various ways:

First, EVs can serve as a resource for demand side management, e.g. demand response (DR) services, as well as for provision of regulation services to the bulk grid, e.g. supply-demand mismatch balancing, especially due to the high penetration of RE sources into energy grids. With hydrogen generation and conversion technologies, FCVs and aerial FCVs will bring additional possibilities to the grid, since reversible FCs using electrolyzers can be employed to derive hydrogen from water. Thus, FCs have been used for power systems in underwater vehicles [38]. It has recently been shown that aerial FCVs such as FC-powered drones have much longer flying times than those equipped with batteries [1]. As such, aerial FCVs have great potential, not only for personal transportation but also for providing services to energy grids, e.g. for responsive DR or to meet emergency needs for energy balancing. Conversely, to extend the flying ranges of FEVs without utilizing charging stations or battery or FC swapping, WPT proposes a solution in which wireless charging can take place along electricity distribution wires, substations, or other suitable places where wireless energy transmitters can be located. An illustration for WPT between ground and flying EVs and smart homes and other grid facilities is depicted in Figure 3.

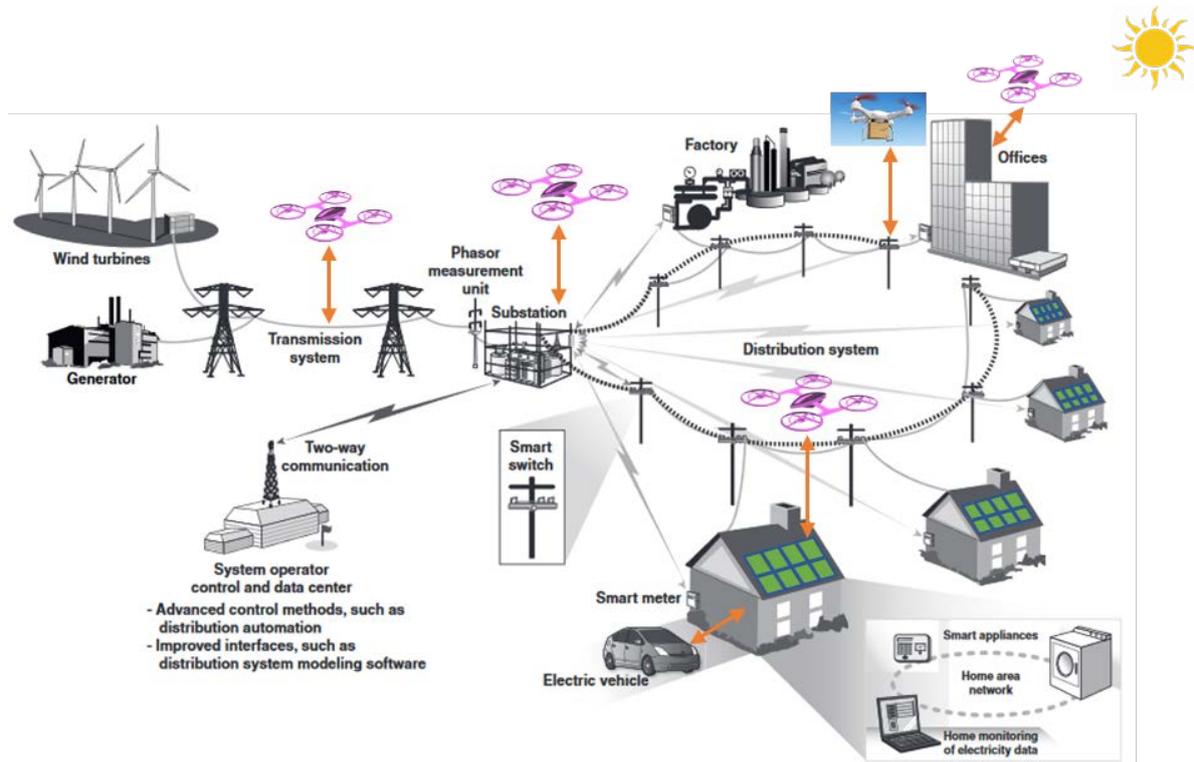

Figure 3. Demonstration for WPT between EVs and energy grid facilities (Adapted from [69]).

The second contribution envisaged from the proposed universal WPT approach is through EVs role in reducing the burden on grid infrastructure through the reduction of the number of fixed charging points. On-the-fly WPT between EVs reduces dependency on energy grids for charging and discharging. Further, the much shorter travel time of FEVs and aerial FCVs allows very short response times to suppress energy mismatches through the bidirectional WPT between FEVs and aerial FCVs and the grid. As an example, Figure 3 also depicts WPT between FEVs and existing grid infrastructure such as electricity distribution poles.

Further, the inspection and maintenance of electric power grids along transmission and distribution lines has been promoted as a potential role for FEVs, especially for locations which are inaccessible to, or dangerous for human workers [49], e.g. for wind turbines, solar panels and high-voltage transmission lines. Further, FEVs can serve as emergency responders for grid accidents such as fire extinguishers for malfunctioning wind turbines inaccessible to regular fire engines.

Finally, renewable energy harvesting infrastructure, e.g. solar panels, on-shore off-shore wind turbines, and sea-wave energy harvester, etc., should be optimally deployed at WPT facilities (e.g. wireless charging bases/stations) for the clean and wireless charging of different types of terrestrial, aerial, and maritime vehicles [60], [62].

## 4.2. Benefits for Transportation Systems

To enable WPT for multiple EVs, in addition to the technical challenges for each specific WPT technology, the motion coordination of EVs must also be contended with. The goal of motion coordination is to keep EVs grouped within a certain range to most efficiently employ WPT technologies for the EV fleet. Depending on the WPT technology employed (i.e. IWPT, CWPT, or OWPT), this challenge may be met in a number of ways. This research takes advantage of a mature methodology known as formation control for coordinating the motion of multiple EVs in parallel. In this method, EVs are controlled (while in motion) to eventually reach and maintain a desired formation, as exhibited using drones as an example in Figure 4.

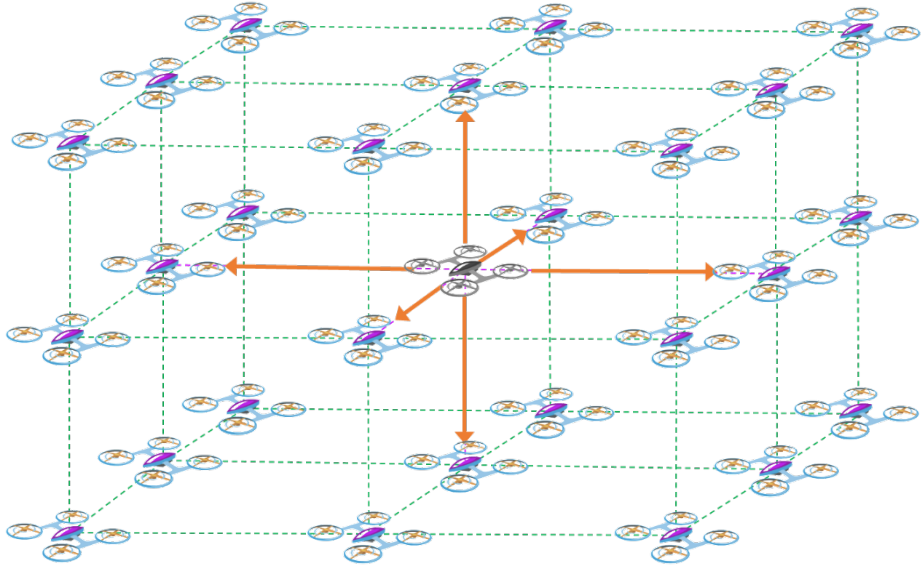

Figure 4. Illustration for lattice formation flying and WPT of multiple FEVs.

The main advantages of moving in formation include energy consumption savings for the vehicle group, and increasing traffic smoothness and throughput, as well as enabling WPT between multiple EVs. Formation movement is most advantageous for autonomous vehicle applications. With FEVs, formation movement is more beneficial since no traffic lights or traffic jams are expected in the sky. Having properly designed sky lanes, the formation motion of FEVs will be extremely efficient, leading to shortened travel times and decreases in energy consumption. Further, sky-based transport has greater potential for longitudinal and transverse lanes than for terrestrial or maritime applications. This approach engenders multiple benefits, including less obstructions for traffic and multiple energy transmission pathways. A recent technological application employing this approach involves Rakuten Ltd., Tokyo Electric Power Company (TEPCO) and Zinrin Ltd. who created a sky highway for delivery drones based on the existing power line layout [42].

The proposed universal WPT concept can help significantly reduce the need for building fixed charging stations or charging points, leading to space saving and a reduction in route searching, traveling time and traveling distance. In the case of flying vehicles these benefits are maximized.

## 4.3. Benefits for the Environment

There is strong evidence that a shift away from fossil fuel-based vehicles (i.e. diesel and gasoline), in preference for battery-based vehicles will yield environmental benefits in the future. Recent research suggests that even EVs which are powered by a conventional fossil-based grid will reap the future benefits of decarbonization at the grid level, unlike their conventional fuel-based counterparts, although challenges remain for the production footprint of EVs [73]. EVs have the potential to reduce both GHGs and particulate matter in the air, leading to not only environmental benefits but also positive health outcomes. On the other hand, waste disposal issues need to be dealt with to ensure that gains made in terms of the environment and health during the use cycle are not reversed at the time of vehicle disposal or during recycling processes [77]. Toward the achievement of national carbon reduction goals, BEVs offer the greatest reduction potential, not only compared to traditional internal combustion engine (ICE) vehicles, but also to recent innovations such as the plug-in hybrid electric vehicle, due to consumer behavior which favor operation in gasoline mode [78]. As nations struggle to meet their carbon reduction targets, more rapid uptake of EVs becomes critical, with evidence emerging from nations with large passenger fleets including China [73], the US [77] and the UK [78].

In terms of environmental impacts of battery production and usage, lithium-ion batteries provide a compelling case as frontrunners for electric vehicle deployment due to their comparatively low emissions and total material requirements (TMR) per energy provided [74]. On the other hand, the use of rare materials in their production means that aggressive recycling will need to be undertaken to ensure supply, or alternative materials will need be identified for future generation batteries [75]. This challenge is not unique to EVs, as alternative fuel vehicles such as hydrogen fuel cell vehicles (FCVs) also use critical materials (notable among them platinum) leading to similar recycling and market impact issues [76].

Should transport shift to a 'sharing economy' approach, gains can be made in terms of overall system cost and environmental improvements, even under current grid constraints [79] – likely to be improved under the in-motion charging elements proposed within this research. Considering elements of human behavior, not only in terms of preferred vehicle type, but also driving patterns, energy system design (such as that proposed in this research) can be undertaken to optimize the energy system and incorporate appropriate energy sources and charging infrastructure so as to maximize environmental gains [80] along with the deployment of complementary technologies at the household level. Building on behavioral aspects, it is anticipated that in addition to grid and charging infrastructure design, that it will be possible to customize consumer EVs to only incorporate the minimum necessary battery quantity, thus delivering lighter, more energy efficient vehicles which are fit for purpose, conserve resources, and fully utilize the integrated charging grid.

### 4.4. Benefits for Convenience

Building on the potential positive impacts which are expected to be engendered for the environment, lifestyle convenience can also be improved through the emergence of a wireless automobile charging network. First and foremost, participation in such a scheme would remain optional, i.e. those who choose to use specific vehicles (i.e. diesel or gasoline hybrid) could do so, and opt-out of the proposed system, their lifestyle would not be impacted. In the same way, users who chose to only charge and discharge their vehicles through privately owned charging infrastructure could also do so, incurring no impact on their current lifestyle.

For those who opt-in to the wireless charging system, several benefits can be obtained, in terms of convenience. First, users of the wireless charging network would not need to stop to charge their vehicle while using it on wireless charging equipped roadways, which we envisage will be predominantly on highways in the first instance. This means that long distance travel, particularly for freight or automated driving systems can become uninterrupted, leading to efficiency gains and reduction in overall transit times. Where systems move toward automated transportation, we also anticipate that traffic jams will be reduced or eliminated as each agent in the network is aware of its peers. Second, it is estimated that eliminating the driver from an electric vehicle, even when the overall cost of the electric vehicle is higher than a traditional alternative, that per ride costs can be reduced by up to 30% compared to driving one's own vehicle, post-2030 [81]. As long as the service provided by automated vehicles is on-par with current chauffeured services (i.e. taxis, UBER etc.) it is anticipated that a significant portion of the population, led by younger working individuals will take advantage of such a system [82]. Where ridesharing is incorporated, lifestyle advantages increase in terms of a cleaner environment and reduced traffic delays, as well as the potential option to forego personal car ownership, reaping a lower overall cost, so long as daily transport needs were predominantly for commuting [83]. In addition to personal lifestyle benefits outlined above, the introduction of autonomous ride sharing options including taxis and buses will improve overall accessibility to reliable transportation options for a broad range of stakeholders, particularly non car owners [83].

## 5. CONCLUSIONS AND FUTURE PATHS

### 5.1. Summary

This paper has identified gaps in existing research regarding WPT, in terms of energy security, availability and convenience, identifying the limited number of wireless charging stations and current WPT technology shortcomings as the main bottlenecks. To mitigate these shortcomings, this research proposes a novel concept of universal WPT using the existing suite of WPT technologies (i.e. IWPT, CWPT, OWPT, etc.) for electrified vehicles as well as robots, electronic devices, and related facilities, whether stationary or in motion. The positive impacts of the proposed concept have been detailed, suggesting multiple benefits across energy and transportation systems, toward the environment in terms of greater deployment of renewable energy, and in terms of convenience and human comfort. Market mechanisms for realizing such a concept have also been defined, leveraging not only the existing grid and additional renewable energy deployment but also distributed ledger technologies, e.g. blockchain, to guarantee the security and privacy of energy transactions. It is anticipated that the realization of the proposed WPT-based energy system concept will significantly contribute to developing low-carbon societies with highly interconnected infrastructure and services, while also improving the comfort and convenience transportation with strong co-benefits to the environment and lifestyle convenience.

### 5.2. Limitations of The Proposed Concept

Limitations of the proposed universal WPT concept mainly arise from WPT technology shortcomings. First, the low efficiency of current WPT systems can impair their deployment for EV usage. While resonant inductive and capacitive WPT technologies have demonstrated high efficiencies, they usually employ a fixed charging pad to wirelessly transfer energy to an EV. Hence, more advanced technologies may be needed in order to realize WPT between moving objects. Moreover, a short distance is also often required to realize highly efficient energy transfer. These short power transmission distances raise concerns regarding the safety of WPT between moving EVs where a safe inter-vehicle distance (when traveling) is speed-dependent and often in the scale of meters. On the other hand, WPT technologies such as OWPT [43], ultrasonic WPT [44] would allow wireless charging at higher distances (i.e.

several meters and beyond) but currently suffer from relatively low efficiency. Microwave WPT using rectennas, RF WPT (e.g. [45–47]), and laser can achieve high efficiency comparable to that of resonant inductive WPT, however capital costs are significantly higher.

The second barrier for large scale WPT deployment is related to safety, in particular the side effects on the EV owner's health or on nearby users due to the high voltage used in capacitive WPT, or the hazards associated with laser and microwave WPT methods.

### 5.3. Recommendations for Future Research

Further theoretical and technical progression is required for universal WPT systems to improve efficiency and user safety, while decreasing capital costs so that universal WPT systems can be widely accepted and deployed.

As analyzed in Section 4, FEVs could bring significant benefits to transportation systems and also to energy systems when the proposed universal WPT concept is utilized. Therefore, a strong push toward the development and deployment of FEVs may provide the best return to society, the environment and people's lifestyle.

Recently, superconducting materials have received attention for their potential use in different types of vehicles due to advantages such as size and weight reduction due to zero conducting resistance at specific operating temperatures. For instance, their use in combination with resonant IWPT for the MAGLEV trains was evaluated in [67]. Another notable use of superconducting materials is on aerial vehicles to achieve light weighting and size reductions. Superconducting motors and hybrid storage appear to be a complementary combination of distributed propulsion and storage approaches.

Artificial Intelligence (AI) including Machine Learning (ML) can significantly help level up the precision, efficiency, and safety of WPT systems. For example, AI approaches can be developed for recognizing charging objects to precisely align transmitters and receivers, detecting other objects in line-of-sight, etc. Recently, attention has been given to foreign object detection for IWPT [66], providing some useful insights for further WPT advancement utilizing AI and ML in the future.